\def \cc{~$\rm{cm}^{-3}$}
\def \msyr{~$M_{\sun}~{\rm yr}^{-1}$}

\def \s{~\rm{s}}
\def \km{~\rm{km}}
\def \kms{~\rm{km}~{\rm s}$^{-1}$}
\def \g{~\rm{g}}
\newcommand{\ud}{\mathrm{d}}

\documentclass{aa}
\usepackage{epsf}
\usepackage{graphicx}
\begin{document}
   \title{Eruptions of the V838~Mon type: \\
          stellar merger versus nuclear outburst models}

\author{R. Tylenda\inst{1} \and N. Soker\inst{1,2}}

\offprints{R. Tylenda}
\institute{Department for Astrophysics, N. Copernicus Astronomical Center,
Rabia\'nska 8, 87-100 Toru\'n, Poland\\ \email{tylenda@ncac.torun.pl}
\and
Department of Physics, Technion-Israel Institute of
Technology,  32000 Haifa, Israel\\ \email{soker@physics.technion.ac.il}}

   \date{Received   }

   \abstract{We discuss various models and scenarios proposed to explain
the nature of the V838~Mon type eruptions. In this class of eruptive objects
we include: M31~RV (erupted in 1988), V4332~Sgr (erupted in 1994) and V838 Mon
(erupted in 2002).
We concentrate on three models: ($i$) thermonuclear runaway on an accreting
white dwarf (nova-like event); 
($ii$) He-shell flash in a post asymptotic giant
branch star
(born-again AGB); and ($iii$) merger of stars. 
We show that models ($i$) and ($ii$) cannot account for the majority
of the observed properties of the objects.
Most significantly, in both nuclear burning type models the object is expected
to heat up before declining and fade as a very hot compact star.
In the observed eruptions the objects declined
as very cool giants or supergiants.
We show that the stellar merger model can account for all the observed properties
and conclude that presently this is the most promising model to explain 
the eruptions of the V838~Mon type.

}

   \titlerunning{Stellar merger model for V838~Mon}

   \maketitle

\section{Introduction \label{int}}

The outburst of V838~Mon, discovered at the beginning of
January 2002 (Brown \cite{brown}), was an important event in the field
of stellar eruptions. Initially thought to be a nova, the object soon appeared
unusual and enigmatic in its behaviour.
The eruption, as observed in the optical, lasted about three months.
After developing an A-F supergiant spectrum
at the optical maximum at the beginning of February~2002, the object
showed a general tendency to evolve to lower effective
temperatures. In April~2002 it almost disappeared from the optical but remained
very bright in infrared, becoming one of the coolest M-type supergiants yet
observed. 

The interest raised by V838~Mon recalled
two similarly behaving objects observed in the past. 
These were M31~RV, a red variable discovered
in M31 in September~1988 (Rich et~al. \cite{rich}), and V4332~Sgr,
discovered at the end of February~1994 (Hayashi et~al. \cite{hayashi}).
Forgotten for almost 10 years
V4332~Sgr appeared, with new observations done in 2003, 
to show very unique spectral features: a strong emission-line spectrum of
very low excitation and a huge infrared excess.

We call M31~RV, V4332~Sgr and V838~Mon
V838~Mon type objects. Their nature is unclear.
As mentioned in a number of papers, thermonuclear models
(classical nova, He-shell flash) seem to be unable to explain
these eruptions. This view is, however, not a consensus opinion
and thermonuclear models are still discussed in the literature.
Munari et~al. (\cite{muna05}) propose
a thermonuclear shell event in an evolved massive star approaching
carbon ignition, while Lawlor (\cite{law}) considers a very late
He-shell flash in a post-AGB (post asymptotic gaint branch) star 
with a period of accretion added during the flash. 
Other proposed explanations invoke stellar
mergers (Soker \& Tylenda \cite{soktyl}) and giants swallowing planets
(Retter \& Marom \cite{ret1}). 

In the present paper we make a comprehensive comparison between various models
and scenarios proposed to explain the nature of the V838~Mon type eruptions
and the observations of these objects. For this purpose, in Sect.~\ref{observ},
we summarize principle observational properties of the objects.
In Sect.~\ref{nuclear}, we discuss thermonuclear eruption models and
we show that they cannot consistently explain the observations.
In Sect.~\ref{merger} we develop the stellar merger model proposed
in Soker \& Tylenda (\cite{soktyl}) and show that it can satisfactorily
account for all the observed properties of the V838~Mon type objects.
A summary is given in Sect.~\ref{summary}.

\section{Observed properties  \label{observ}}

The principal observational property that makes M31~RV, V4332~Sgr and V838~Mon
form a separate class of eruptive objects is their spectral and photometric
evolution in late phases of eruption, in particular during decline. 
The evolution during outburst has been observed only in V838~Mon, so we do
not know if its light curve properties, i.e. multi-outburst character
(Crause et~al. \cite{lisa03}, Retter \& Marom \cite{ret1}),
are also characteristic of the other two objects. 
Both M31~RV and V4332~Sgr very likely had been
discovered at a late phase of eruption. Their observed evolution was very
similar to that of V838~Mon starting in last days of March~2002, i.e. about
3 months after its discovery.
In this phase the objects resambled K-type or early M-type
supergiants and on a time scale of days they evolved to later
spectral types (Mould et~al. \cite{mould}, Martini et~al. \cite{martini},
Tylenda \cite{tyl05}).
Soon they became the coolest known supergiants or giants.
The latest recorded spectrum of M31~RV was M6\,I, corresponding to 
$T_\mathrm{eff} \simeq 2600$~K (Mould et~al. \cite{mould}), but after, 
judging from the subsequent infrared photometry, 
the object evolved to even lower effective temperatures.
V4332~Sgr reached a spectral type as late as M8-9\,III and 
a corresponding effective temperature of $\sim 2300$~K 
(Martini et~al. \cite{martini}), although it is not
excluded that the object dropped to even lower $T_\mathrm{eff}$.
In the coolest
phase of V838~Mon its spectral type was usually being refered to as later
than M10 (Desidera \& Munari \cite{desmun}) 
while its effective temperature was as low as $\sim 2000$~K (Tylenda \cite{tyl05}).

Before reaching early M-type spectral classes, i.e. for $T_\mathrm{eff} \ga 3600$~K,
the objects evolved at a roughly constant luminosity. Near the spectral type
M0 all the three objects began to decline. Initially
the luminosity decrease was rather fast and a factor
of 10 in the luminosity drop was reached after $\sim 50$, $\sim 10$ 
and 50--100~days in the case of M31~RV (Mould et~al. \cite{mould}),
V4332~Sgr (Tylenda et~al. \cite{tcgs}) and V838~Mon (Tylenda \cite{tyl05}), 
respectively. At this luminosity level V838~Mon and
presumably also V4332~Sgr were
close to their minimum effective temperatures. All that is known about 
the subsequent evolution of M31~RV is that it became fainter by a factor of 100
after $\sim 300$~days (Mould et~al. \cite{mould}).
V838~Mon, after having reached its coolest state, kept declining
in luminosity but at a much slower rate and with a slowly increasing
effective temperature. About 1000~days after eruption the object was 
$\sim 40$ times fainter than at maximum, with $T_\mathrm{eff} \simeq 2600$~K
corresponding to a spectral type of $\sim\ $M6 (Tylenda \cite{tyl05}).
At $\sim 3500$~days after eruption 
V4332~Sgr decreased by a factor of $\sim 1500$ in luminosity, its spectral
type was then $\sim\ $M3 and $T_\mathrm{eff} \simeq 3400$~K 
(Tylenda et~al. \cite{tcgs}). For the three objects there were
no indications of any excursion to high effective temperatures during decline
and no nebular stage was observed. The objects declined at very low
effective temperatures. 

The maximum luminosity attained by an eruptive star is an important observational
parameter in understanding the nature of the object. Unfortunately it depends
on the distance which is usually uncertain for Galactic objects. Therefore
M31~RV is the only object for which we can be confident of
its maximum luminosity of $\sim 8 \times 10^5\ L_{\sun}$ 
(Mould et~al. \cite{mould}).
For a distance to V838~Mon of 8~kpc the maximum luminosity of this object
was $\sim 1.2 \times 10^6\ L_{\sun}$ (Tylenda \cite{tyl05}). 
However, because of distance uncertainties
this value is probably uncertain by a factor of $\sim 2$. 
Even more uncertain is the
maximum luminosity of V4332~Sgr. If its progenitor had been a main sequence
star the object would have attained $\sim 5 \times 10^3\ L_{\sun}$ 
(Tylenda et al. \cite{tcgs}),
which would be significantly lower than in the two other cases. 

The outburst amplitude is independent of distance so, in principle, it can be 
determined more precisely than the luminosity. Unfortunately only an upper limit
is available for the brightness of M31~RV before outburst so it can only be
said that the object brightened by at least $\sim 5$~mag in the $I$ band
(Mould et~al. \cite{mould}).
More data are available for the progenitors of V4332~Sgr and V838~Mon so it
can be estimated that at maximum these objects were, respectively, $\sim 5 \times 10^3$ and
$(0.5 - 1.5) \times 10^3$ times more luminous than before the eruption
(Tylenda et~al. \cite{tcgs}, \cite{tss}, Tylenda \cite{tyl05}).

The three objects probably lost significant mass. However, direct evidence,
i.e. P-Cygni line profiles, have been observed only for V838~Mon. 
They give wind velocities
up to $\sim 600$\kms\ although most of mass loss occured at $150 - 400$\kms
(Munari et~al. \cite{munari}, Kolev et~al. \cite{kolev}, 
Crause et~al. \cite{lisa03}, Kipper et~al. \cite{kipper}).
Similar values, i.e. $270 - 400$\kms, result from the observed expansion
of the effective photosphere in eruption (Tylenda \cite{tyl05}).
Widths of emission lines observed in V4332~Sgr, if interpreted as primarily due to
expansion, indicate a velocity of $\sim 100$\kms (Martini et~al. \cite{martini}).
Expansion velocities of a few hundreds\kms\ have been estimated for M31~RV 
(Mould et~al \cite{mould}, Iben \& Tutukov \cite{ibentutu}).

The total mass lost in eruption is uncertain for the objects.
Very rough estimates done for M31~RV give
values between $10^{-3} - 10^{-1}\ M_{\sun}$ (Mould et al. \cite{mould}).
No estimate has been done for V4332~Sgr. 
Rushton et~al. (\cite{rush03}) obtained an upper limit of
$\sim 0.01\ M_{\sun}$ for the total mass lost from V838~Mon 
from their CO and SiO observations. This upper limit
should, however, be multiplied by at least a factor of 10.
Rushton et~al. have integrated their upper limits for
the line fluxes over a velocity interval of 50\kms, 
which is much too low compared to the observed expansion velocities (see above).
Also, their estimates
assume that all the ejecta were seen in the lines (optically thin limit).
Tylenda (\cite{tyl05}) concludes that the total mass lost by V838~Mon is between
$\sim 5 \times 10^{-3}\,M_{\sun}$ and $\sim 0.6\,M_{\sun}$. Lynch et~al. 
(\cite{lynch}) give $\sim 0.04\,M_{\sun}$ with a large uncertainty. 
An integration of the luminosity derived by Tylenda (\cite{tyl05}) over 
the pre-eruption and eruption phase gives a value of $\sim 2.5 \times 10^{46}$~ergs. 
Assuming equipartition of energy between radiation and mass loss, as well as
a wind velocity of 300\kms\ one obtains mass loss of $\sim 0.03\,M_{\sun}$.

In the case of V4332~Sgr and V838~Mon inverse P-Cygni profiles were observed
during the decline, indicating an infall velocity of $\sim 20$\kms 
(Martini et~al. \cite{martini}, Rushton et~al. \cite{rush05a}).

The three objects in decline show strong bands from molecules involving oxygen,
clearly indicating oxygen-rich (C/O~$<$~1) matter. More detailed determinations
of the chemical composition are available only for V838~Mon 
(Kipper et~al. \cite{kipper}, Kaminsky \& Pavlenko \cite{kam}).
The abundances
of the iron group metals are slightly below the solar values, 
$\mbox{[Fe/H]} \simeq -0.3$. Those of Li, Ba and La are somewhat above, 
although as noted in Kipper et~al. (\cite{kipper}), the obtained abundances
of these elements are sensitive to the (uncertain) microturbulence
velocity. Thus it can be concluded that the observed abundances in V838~Mon
are consistent with unprocessed
matter at the galactocentric distance of the object.

Nothing is known about the progenitor of M31~RV. In the case of V4332~Sgr, archival
photometric data suggest an F--K type star (Tylenda et~al. \cite{tcgs},
Kimeswenger \cite{kimes05}).
V838~Mon has a B3\,V spectroscopic companion (Munari et~al. \cite{mundes}).
The companion is very likely to be related to V838~Mon itself, e.g. forming a binary system,
although a pure coincidence in the sky cannot be excluded. If it is a binary, V838~Mon
before the outburst was a star similar to its B-type companion, i.e. an early B-type
main-sequence star or an A-type pre-main-sequence star (Tylenda et~al. \cite{tss}).
The object is surrounded by extended circumstellar matter seen in the light echo
(Munari et~al. \cite{munari}, Bond et~al. \cite{bond}). 
The matter is most likely of interstellar
origin, possibly being part of one of the star forming regions seen in the direction of
and at similar distances as V838~Mon 
(Tylenda \cite{tyl04}, Tylenda et~al. \cite{tss}).

Circumstellar matter is also seen in V4332~Sgr, strongly radiating in the infrared as
observed $\sim 9$~years after outburst (Banerjee et~al. \cite{bval}, \cite{bva}).
Most likely this matter forms a circumstellar
disc dissipating its energy in viscous processes (Tylenda et~al. \cite{tcgs}).
The origin of the disc is however
unclear. The object also shows an intense emission-line spectrum
due to neutral elements and molecules (Banerjee \& Ashok \cite{banash},
Tylenda et~al. \cite{tcgs}).
A possible interpretation is that this emission
is due to optically thin (in the continuum) matter left after outburst and 
now orbiting the central star at a few stellar radii (Tylenda et~al. \cite{tcgs}).

V838~Mon was observed with Chandra a year after outburst 
(Orio et~al. \cite{orio03}). The object was not detected and
an upper limit to the X-ray luminosity was $\sim 0.13\ L_{\sun}$ for a distance
of 8~kpc.

SiO maser emission was detected in V838~Mon $\sim$~3~years after its outburst
(Deguchi et~al. \cite{deguchi}). Its intensity increases with time (Claussen
et~al. \cite{clauss}).

\section{Problems with thermonuclear eruption models \label{nuclear}}

A sudden brightening of a star by a factor of $\sim 10^3$ is usually
interpreted as due to a thermonuclear eruption. This is because
most astrophysical events of this range are nova-type outbursts.
Their observational properties can be consistently explained
by thermonuclear runaway on an accreting white dwarf. Another 
mechanism based on nuclear burning
is a He-shell flash
in an evolved star. Here we discuss the observed
properties of the V838~Mon type eruptions in terms of these two
thermonuclear mechanisms. 

\subsection{Nova-like outbursts  \label{nova}}

The V838~Mon type eruptions may be considered as nova-like events.
The nova mechanism, thermonuclear runaway on an accreting white dwarf, 
has been well studied and is known to be able to account for a large variety
of astrophysical outbursts. This includes relatively "gentle" 
eruptions lasting decades in some symbiotic stars, 
slow novae that require months to reach maximum
and that then can stay active for years, and fast novae which in an hour or so
increase their brightness by 3--4 orders of magnitude and violently eject
matter with velocities well above 1000\kms. Indeed several objects of the nova type,
like some slow novae, have light curves similar to that of V838~Mon.
An example is HR Del (Nova Del 1967), which showed an initial increase in
luminosity that lasted $\sim 5$ months, then two luminosity peaks
with a five month separation between them, and then a decline over many
years (e.g., Drechsel et~al.\ \cite{drec}; Rafanelli \& Rosino \cite{rafa}).
Scaling the luminosity upward and shrinking the time scale by a
factor of $\sim 5$ will lead to a light curve qualitatively
similar to that of V838~Mon.
Friedjung (\cite{fried}) pointed out the unusual behavior of
this nova in having an almost stationary photosphere.
This is qualitatively similar to V838 Mon, which
also seemed to have a quasi-static photosphere in its pre-outburst
(January~2002) phase (Tylenda \cite{tyl05}).
However, the light curve is probably the only observational
feature of the V838~Mon type objects that can be easily interpreted 
in terms of a nova-like event.
As already noted in early analyses of M31~RV
(Mould et~al. \cite{mould}) and V4332~Sgr (Martini et~al. \cite{martini}), 
several observational properties of these objects
are difficult to reconcile with the nova hypothesis.

The crucial argument against the nova-like event
comes from the observed spectral evolution 
of the objects during outburst and decline.
In the case of a classical nova event, after an outburst on a degenerate white dwarf, 
steady nuclear burning commences, which halts the luminosity typically 
at $\sim 10^4\ L_{\sun}$. It can be extinguished only after the envelope mass
becomes too small to keep the burning shell dense and hot enough.
But this implies that the photospheric radius
must be small, hence the effective temperature high.
This is the reason why novae after eruption (but also post-AGB stars and 
late helium shell flash -- born-again AGB -- objects, 
discussed in Sect.~\ref{he_flash}) heat up 
while evolving at late phases. Models, as well as observations, show that
$T_\mathrm{eff}$ of novae can then reach values well above $10^5$~K, giving rise
to a nebular stage and a significant flux in X-rays.
When the nuclear burning stops, the object
decreases in luminosity as a very {\it hot} star.
In the H--R diagram, the three objects, i.e. RV~M31, V4332~Sgr and V838~Mon,
evolved in the {\it opposite} direction to a nova. 
As discussed in Sect.~\ref{observ}, they evolved to lower
$T_\mathrm{eff}$ and faded as very {\it cool} stars.
So far the largest luminosity decrease has been recorded for V4332~Sgr,
which faded by a factor of $\sim 1500$ in 9~years, still being an M-type
object. In our opinion, this type of evolution in luminosity and 
effective temperature is evidence that the V838~Mon type eruptions
are {\it not} nove-like events. Below we discuss this and other points
concerning the nova mechanism in more detail and in relation to individual objects.

Iben \& Tutukov (\cite{ibentutu}) have proposed a nova-like scenario to account for
the M31~RV eruption. They consider a low-mass ($\sim 0.6\ M_{\sun}$) 
cold white dwarf 
in a short-period cataclysmic binary, which accretes at a very low rate
($\sim 10^{-11}\ M_{\sun}\ \mathrm{yr}^{-1}$). 
This allows it to accumulate a relatively large
mass ($\sim 0.005\ M_{\sun}$) before thermonuclear runaway.
The authors argue that in this case the nova outburst would be energetic 
(maximum luminosity well above the Eddington limit) 
and the object can evolve to low effective temperatures
due to a massive expanding envelope remaining optically thick for a few years.
This scenario does not allow one to avoid the hot phase
after the outburst. Sooner or later the expanding envelope becomes transparent
and the steady burning white dwarf would have to be visible. The phase of steady burning 
is very long for low mass white dwarfs. From the results of Yaron et~al. (\cite{yaro})
it can be inferred that for an $0.65 M_{\sun}$ white dwarf it lasts at least
$\sim 100$~yrs. The scenario also implies that the binary system is old
($\sim 10^{10}$~yrs) and that the main sequence companion is of low mass 
($\la 0.3\ M_{\sun}$), i.e. an M type dwarf. 

In the case of M31~RV the scenario of Iben \& Tutukov cannot
be conclusively confirmed nor definitively rejected. 
We know nothing about its progenitor. We have no data on the object
for epochs later than $\sim 100$~days after its discovery in eruption,
so a hot phase, when the object is expected to be quite faint in optical,
cannot, in principle, be excluded. 
However, the hot phase is likely to produce strong
emission lines when the expanding matter becomes ionized. It seems
rather unlikely that a nebular stage of M31~RV escaped detection in
spite of numerous surveys for planetary nebulae done for M31 in the last decade.

In the case of V4332~Sgr the scenario of Iben \& Tutukov cannot be applied.
The progenitor was most probably an F--K type star.
Thus, if identified as a donor secondary, it would
imply a system in the upper range of the secondary mass 
($\sim 1\ M_{\sun}$) of cataclysmic binaries.
However, the $B-R$ colour from archive data discussed in
Tylenda et~al. (\cite{tcgs}) is uncertain so an M type progenitor,
required by Iben \& Tutukov, cannot
be definitively excluded. But then its absolute magnitude would be
$M_V \ga +9.0$, the distance to the object would be $\la 300$~pc
and the luminosity during outburst would be $\la 150\ L_{\sun}$.
Such a low luminosity is obviously ruled out in any white dwarf 
thermonuclear event.

Thus, if continuing with the nova-type scenario in the case of V4332~Sgr, 
an $\sim 1 M_{\sun}$ secondary has to be accepted. 
This however implies a typical mass transfer rate
($\gg 10^{-11}\ M_{\sun}\ \mathrm{yr}^{-1}$) and a typical nova outburst.
Then there is no way to evolve to M spectral types 
after the outburst maximum. Second, there is no way to avoid the
nebular stage. The object is now $\sim 1500$ times fainter than at maximum
so the white dwarf had to switch off nuclear burning and thus it had to have gone
through a very hot phase in the past giving rise to a nebular stage. 
Although we have almost
no data on the object between 1995--2002, it seems rather unlikely that its
nebular stage would have escaped detection in spite of numerous surveys for emission-line
objects (e.g. planetary nebulae) done in the last decade in the Galactic disc
and bulge (galactic coordinates of V4332~Sgr are $l = 13\fdg63,
b = -9\fdg40$). A luminosity drop by a factor of 1000 on a time scale 
of a few years (note that the initial drop in lumonosity by a factor of 100
occured over 3 months, Martini et~al. \cite{martini}) requires a massive
white dwarf in the nova model, which in turn is expected to result in 
expansion velocities $\ga 1000$\kms (e.g. Yaron et~al. \cite{yaro}).
This is at least an order of magnitude higher than the value of $\sim 100$\kms
observed in V4332~Sgr. 
Finally, the observed radius of the V4332~Sgr
remnant in 2003, i.e. $\sim 5\ R_{\sun}$ compared to $\sim 1\ R_{\sun}$
for the progenitor (Tylenda et~al. \cite{tcgs}),
implies that the whole (hypothetical) nova-like binary would still be 
embedded in a common envelope. 
Then how was the white dwarf
able to extinguish nuclear burning if it still has an extended hydrogen-rich
envelope?

The progenitor of V838~Mon was a rather hot and luminous star. If the object
forms a (presumably very wide) binary with the B3\,V companion identified
by Munari et al. (\cite{mundes}), then the progenitor was either a B2--3
main sequence star with a mass $\sim 8-10\ M_{\sun}$
or a pre-main-sequence star of $\sim 5\ M_{\sun}$ (Tylenda et~al. \cite{tss}).
In the former case, one can propose evolutionary paths in binaries
in which a white dwarf can be
formed from an initially more massive companion, especially if
mass transfer took place between the binary components. 
However if the cooling time ($\ga 10^7$~years) necessary
for the white dwarf to cease emitting significant ionizing flux (no significant
ionization is observed in the diffuse matter surrounding V838~Mon), as well as
to become able to produce a nova-type outburst,
is taken into account, very little room remains for this possibility.
The scenario seems to be very unlikely given the fact that we do not
know of any cataclysmic variable having a B-type main sequence secondary.
In the former case the situation is even worse. The age of a $\sim 5\ M_{\sun}$
pre-main-sequence star can be estimated as $\sim 5 \times 10^5$~years.
There is no way to produce and cool a white dwarf companion in such a
short time.

If V838~Mon and the B3\,V companion are observed at the same position in the sky
due to a random coincidence then, apart from the above possibilities, the progenitor
of V838~Mon could have been either a BA-type giant
of mass $\ga 2.5\ M_{\sun}$ or a B-type post-AGB star (Tylenda et~al. \cite{tss}). 
In both cases
a white dwarf companion cannot be excluded. In the former case the giant
would be in a rapid expansion phase. Thus if mass transfer begins it would
increase rapidly to rates close to, or even above, the Eddington limit 
for accretion on a white dwarf. The white dwarf would thus be bright and 
the whole system would remind a symbiotic star. In these conditions
the hydrogen burning on the white dwarf surface
would be ignited rather gently leading to an increase in the system  brightness only
by a factor of a few (compared to the giant luminosity plus 
the white dwarf accretion luminosity). V838~Mon increased in luminosity
by a factor of $\sim 1000$.

The case of a B-type post-AGB star implies that a few thousand years ago the
object was an AGB giant. Thus the only possibility to get a nova-type
event is that the white dwarf companion was accreting from the AGB wind.
However, in order to reach a nova peak luminosity above $10^5\ L_{\sun}$
the accretion rate would have to have been $\la 10^{-10}$\msyr 
(Yaron et~al. \cite{yaro}). Taking 
a typical AGB mass loss rate of $10^{-6}$\msyr (note that at the end of AGB,
i.e. just before the post-AGB phase, mass loss is expected to be significantly
higher than that) a wind velocity of
10\kms\ and using Eq.~(4.38) in Frank et~al. (\cite{fkr})
one finds that the above accretion rate onto a $1\ M_{\sun}$ white dwarf
is possible if the binary separation is $\ga 2 \times 10^5\ R_{\sun}$.
Such a large binary separation is excluded. It would be $\ga 100$ times 
larger than the maximum effective radius
of V838~Mon during eruption (Tylenda \cite{tyl05}).
Thus the post-AGB component would be visible all the time, including
the decline phase. Yet the progenitor was not seen in the decline phase
when V838~Mon became very cool and invisible in the optical (e.g. Munari
et~al. \cite{muna05}).

Thermonuclear outbursts can in principle supply a huge amount of energy
(e.g., Sparks et al. \cite{spar})
with the maximum luminosity
at peak of $\ga 10^{5}\ L_{\sun}$ achieved by fast novae.
In V838 Mon the maximum luminosity of $\sim 10^6\ L_{\sun}$ lasted $\sim 70~$days.
Novae with a decline of 2 magnitudes in $t_2\sim 100~$days,
have for their maximum visual luminosity
$L_V({\rm max}) < 10^5\ L_{\sun}$, and more typically
$L_V({\rm max}) \simeq 2 \times 10^4\ L_{\sun}$ (Cohen \cite{cohen}).

The outflow velocity observed in V838 Mon, typically
$\sim 150-400$\kms,
can be matched with slow novae, but not with fast novae.
Some models calculated by Yaron et al.\ (\cite{yaro}) reach a nova peak
luminosity of $\sim 10^6 L_{\sun}$. However, the ejection speeds in these
cases are $>2000$\kms, much faster than observed in V838~Mon.

All the nova models in the extensive grid of Yaron et~al. (\cite{yaro})
eject $< 10^{-3}\ M_{\sun}$. In the case of the most luminous ones, i.e.
having a maximum luminosity $\sim 10^6 L_{\sun}$, it is even 
$\la 5 \times 10^{-5}\ M_{\sun}$. Estimates for V838~Mon
give mass loss $> 5 \times 10^{-3}\ M_{\sun}$, most likely
$\sim 0.01 - 0.1\ M_{\sun}$. Simliar mass loss also probably occured
in M31~RV.

V838~Mon is embedded in diffuse matter which resulted in the spectacular light
echo following the outburst. The light echo phenomenon is not 
a common feature of novae. This is what is expected. Fast (previous) ejecta
should sweep up any circumstellar or interstellar matter from the vicinity
of the nova.
The light echo has been observed only in the case of 
the fast nova GK~Per (Nova Per 1901)
(Ritchey \cite{ritch}, Couderc \cite{coud}).
The echoing dust in GK~Per was either
interstellar in origin (Hessman \cite{hess}) or due to an
old planetary nebula (Bode et~al. \cite{bode}).
Tylenda (\cite{tyl04}) and Tylenda et~al. (\cite{tss}) argue that the echoing
matter in V838~Mon is interstellar. This is compatible with the idea that
V838~Mon is a young object rather than an evolved nova-like system.

The strong infrared excess observed now in V4332~Sgr strongly suggests
that the object is surrounded by a large (inner radius of $\sim 30$ 
central star radii) and luminous
($\sim 15$ times brighter than the central star) accretion disc 
(Tylenda et~al. \cite{tcgs}). It is unlikely
that a structure like this
could have formed arround a nova-type binary.

The abundances usually are important in identifying the nature of an object. 
Simulations show that novae with
very large peak luminosity are expected to eject highly processed matter
with $Z$ up to $\sim 0.5$ (Yaron et~al. \cite{yaro}). 
As discussed in Sect.~\ref{observ}, the observed abundances in V838~Mon 
are compatible with unprocessed matter.
Although there are no data on
the abundances of key elements for novae,
such as C, N, O, Mg and Ne, no strong molecular CN lines were
observed in V838~Mon; strong CN lines are expected from
novae at early stages (Pontefract \& Rawlings \cite{pont}).

As noted in Orio et~al. (\cite{orio03}), the lack of measurable
X-ray flux in V838~Mon is evidence against a nova-like runaway in this object.

\subsection{Helium shell flashes  \label{he_flash} }

While in novae thermonuclear runaway occurs on a relatively
cool white dwarf, in He-shell flashes the runaway process takes place
in a shell at the outskirts of the core of
a nuclear active star.
This takes place in AGB and post-AGB stars. 
Before the flash the energy generation is maintained 
by H-burning
that adds more helium to the He-shell, until conditions
for ignition are reached.
One of the characteristics of this process is that because of
expansion and cooling, the H-burning is extinguished as
the He-shell ignites. As a result, the surface luminosity typically varies
only by a factor of 2--3 during the thermal pulse (Wood \& Zarro \cite{wood},
Vassiliadis \& Wood \cite{vass}, Herwig \cite{herw}).
Thus there is
no way to obtain an increase of $\sim 10^3$ in luminosity, as observed
in V838~Mon and V4332~Sgr, from a thermally pulsating AGB star.
In the case of V838~Mon the AGB hypothesis has to be rejected also 
on observational evidence. 
As shown in Tylenda et~al. (\cite{tss}) the progenitor of this object
was significantly hotter than a KM-type giant or supergiant.

As discussed in Tylenda et~al. (\cite{tss}) it is possible (although unlikely)
that the V838~Mon progenitor was a B-type post-AGB star. An He-shell
flash, usually called a late He-shell flash, 
can also occur  at this evolutionary stage. However, for the same reason 
as discussed above (an active H-shell) only a moderate (at most a factor of 10)
increase in brightness is possible (Iben \cite{iben84}, Bl\"ocker \& Sch\"onberner
\cite{bs97}).

The only way to get a significant luminosity increase during
a He-shell flash is the so-called very late He-shell flash or
born-again AGB event which occurs
on the cooling part of the post-AGB evolutionary track when the H-shell
is already extinguished (Iben \cite{iben84}, Herwig \cite{herw01}, 
Lawlor \& McDonald \cite{lmacd}). However in this case the object is expected
to be very hot ($\sim 10^5$~K) before the flash and surrounded by a glowing
planetary nebula. Indeed, objects identified as undergoing
a (very) late He-shell flash, i.e FG~Sge, V4334~Sgr, V605~Aql, are surrounded
by planetary nebulae. Nothing of this kind has been reported for the V838~Mon
type objects. In the case of V838~Mon itself we know from its light echo that the
object is surrounded by diffuse matter but this matter is {\it not} ionized
(Orio et~al. \cite{orio02}, Munari et~al. \cite{munari}). For the
same reason (lack of ionization of the light echo matter) the idea
of Munari et~al. ({\cite{muna05}), namely that the eruption of V838~Mon
was due to a thermonuclear shell flash in an evolved massive star 
(initial mass $\sim 65\ M_{\sun}$) close to
the Wolf-Rayet phase ($T_\mathrm{eff} \simeq 5 \times 10^4$~K), has
to be rejected (Tylenda et~al. \cite{tss}).

Another important point concerning the late He-shell flash is that after 
the initial excursion to high luminosities and low effective temperatures,
the object is expected to repeat its previous evolutionary path toward
high effective temperatures. The evolution is thus qualitatively the same as
that of a nova at late stages. As discussed in Sect.~\ref{nova} this type 
of evolution is in a sharp disagreement with the observed evolution of
the V838~Mon type objects. This is one of the crucial arguments against
interpreting these objects as undergoing a (very) late He-shell flash.

Finally, as noted in Kipper et~al. (\cite{kipper}), the abundances
observed in V838~Mon do not support the idea of a late He-shell flash.
In particular all the objects identified as flashing post-AGB stars 
(FG~Sge, V4334~Sgr, V605~Aql, as well as central stars of A\,30
and A\,78) are C-rich while all the V838~Mon type objects have C/O$ < 1$.

\subsubsection{The born-again model of Lawlor  \label{lawlor}}

Lawlor (\cite{law}) proposed a born-again AGB model 
including an episode of accretion to account for the V838~Mon 
eruption. The accretion is supposed to be due to a burst of mass loss from 
a main-sequence binary companion irradiated by the He-shell flashing white dwarf.
The addition of accretion produces a second outburst,
which the author identifies with the main
eruption of V838~Mon that started in the first days of February~2002.

The arguments raised above against the very late He-shell flash
also hold in the case of Lawlor's model. 
This model assumes a very hot progenitor
and must evolve to high effective temperatures before final fading.
Other important problems appear when comparing assumptions and results
of the model with the observations of V838~Mon.

As the main observational test Lawlor (\cite{law}) calculated a visual light
curve from his model and obtained a result qualitatively similar to 
the observed one assuming a distance of 6.3~kpc. However when calculating the model
light curve Lawlor did not take into account interstellar extinction which
in the case of V838~Mon is as large as $A_V \simeq 2.7$ (Munari et~al. \cite{muna05},
Tylenda \cite{tyl05}). Thus an agreement between the model and observed light curves
can be obtained but for a distance as small as 1.8~kpc, which can be compared
to e.g. a lower limit of 5--6~kpc obtained from the light echo expansion
(Bond et~al. \cite{bond}, Tylenda \cite{tyl04}, Tylenda et~al. \cite{tss}).
For more reasonable distance estimates of 8--10~kpc the model of Lawlor is 
systematically too faint by a factor of 20--30.

Another important problem is that
the surface parameters of Lawlor's model just before the accretion
event are $T_\mathrm{eff} \simeq 40\,000$~K and 
$R \simeq 1.5\ R_{\sun}$. This phase is supposed to correspond
to the pre-eruption phase of V838~Mon in January~2002. However, the observed
parameters of V838~Mon during this phase were $T_\mathrm{eff} \simeq 5\,000$~K
and $R \simeq 350\ R_{\sun}$ (Tylenda \cite{tyl05}).
This discrepancy ensures, as we discuss below, that Lawlor's idea
of an irradiation-induced accretion event cannot occur in the case of V838~Mon.

Lawlor (\cite{law}) proposes that the irradiated companion is the B3\,V companion 
discovered by Munari et~al. (\cite{mundes}). However this companion must be at 
a distance of at least $\sim 3000\ R_{\sun}$ from V838~Mon 
(effective radius of V838~Mon at the
epoch when the companion was discovered -- Tylenda \cite{tyl05}) 
and most probably is much father away. Obviously, no important irradiation 
effects nor significant mass exchange can be expected at such a large distance 
between the components.
Thus the only possibility is that the star identified as the progenitor of V838~Mon
in Tylenda et~al. (\cite{tss}) was irradiated by its hypothetical flashing 
white-dwarf binary companion. This star most probably was an early B-type 
main sequence star prior to the V838~Mon eruption.
However, since the irradiation-induced mass transfer is supposed to occur
during the pre-eruption of V838 Mon, i.e. in January 2002,
the B-type star would have to be at a distance of at least $\sim 1600\ R_{\sun}$
from V838~Mon (supposed to be the flashing white dwarf). Otherwise V838~Mon, inflated
to $\sim 350\ R_{\sun}$ in the pre-eruption phase, would fill up
its Roche lobe and mass transfer would occur in the direction opposite to that
required in the model of Lawlor. Obviously the B-type main sequence star would be
deeply inside its Roche lobe, so in order to obtain significant mass loss from it
(Lawlor's model requires accretion at a rate of $\sim 10^{-3}$\msyr) 
the irradiated flux would have to be significantly greater than the
surface flux of the B-type star. This would require
an irradiation source as luminous as $\ga 2 \times 10^8\ L_{\sun}$. 
In January~2002 V838~Mon 
had an effective temperature of $\sim 5000$~K (Tylenda \cite{tyl05}, 
see also e.g. Kimeswenger et~al. \cite{kimes}, Rushton et~al. \cite{rush05b}), 
i.e. it was much cooler than an early B-type star. 
For obvious reasons a cooler object cannot 
significantly heat up a hotter one even if the cooler one is much more luminous.

As discussed in Sect.~\ref{observ} V838~Mon lost $0.01 - 0.1\ M_{\sun}$
of H-rich matter during its eruption. 
There seems to be no way to get such a large mass loss from
the model of Lawlor. The mass of the H-rich envelope of a white dwarf before
a very late He-shell is $< 10^{-3}\ M_{\sun}$ (Iben \cite{iben84}) 
and $< 10^{-4}\ M_{\sun}$
has been accreted in Lawlor's model. Intense mass loss from
Lawlor's model would quickly expose highly processed matter at the surface. 
Neither in the 2002 eruption nor during the decline was there  
any observational indication of this. In particular, the C/O ratio remained
$< 1$ all the time.

We can conclude that the model presented in Lawlor (\cite{law}) cannot be accepted
as an explanation of the V838~Mon eruption. It is discrepant with the observations
in too many points.

\section{The stellar merger model \label{merger}}

In the previous section showed that neither the nova-like runaway
nor the He-shell flash can account for the observed properties of the
V838~Mon type eruptions. The only other way to liberate
a significant amount of energy in stellar astrophysics
is release of gravitational energy, i.e. gravitational contraction or
accretion. Gravitational contraction powers protostars.
It is possible, as discussed in Tylenda et~al. (\cite{tss}), that the progenitor
of V838~Mon was a pre-main-sequence star. 
However there is no known way in which an energy
of $\ga 3 \times 10^{46}\ \mbox{erg}\ \mbox{s}^{-1}$ could be
released on a time scale of a month in a pre-main-sequence star.

Thus we have to consider accretion events. The more compact the accreting object,
the more effective is the accretion in terms of energy liberated per unit
mass accreted.
Accretion onto a neutron star or a black hole, being the most effective,
is however excluded as it would result
in an X-ray source. Accretion onto a white dwarf would have to procede
at a super-Eddington rate to approach the observed luminosities
and would lead to a nova-type thermonuclear event
with all the problems involved, as discussed in Sect.~\ref{nova}.

We are left with accretion on a main sequence star.
In this case, to obtain a luminosity of $\sim 10^6\ L_{\sun}$ 
the accretion rate must be $\ga 0.01$\msyr. Taking into account the 
total energy involved in the V838~Mon event, as estimated in 
Sect.~\ref{merg_sc},
the main sequence star would have to accrete $\ga 0.1\ M_{\sun}$ on 
a time scale of months. An event like this is excluded in 
secular binary system evolution, as there is no mechanism
that could result in the transfer of such a large amount of mass
in such a short time. Young main sequence or pre-main-sequence stars are often 
surrounded by massive protostellar discs. Thermal instabilities in the disc,
similar to those in dwarf novae, can transfer a significant mass from the
disc to the central star. This is the mechanism believed to
produce FU~Ori type eruptions (Hartmann \& Kenyon \cite{hartken}).
The discs in these objects can transfer as much as $0.01\ M_{\sun}$. However
this occurs on a time scale of a hundred years.

Therefore it seems that the only way to get the accretion event required to explain 
the V838~Mon type eruptions is accretion of a low mass object
onto a main sequence star. This is the reason why we proposed
a stellar merger model in Soker \& Tylenda (\cite{soktyl}) to explain
the eruption of V838~Mon. At that time the distance to the object
was believed to be $\sim 1$~kpc. Therefore, the merging
of two low-mass main-sequence stars was considered.

In the present paper the qualitative general scenario for the eruption of V838 Mon
remains similar to that in Soker \& Tylenda (\cite{soktyl}). 
We however change some details and update
numerical values to be consistent with the new distance estimate
and more recent data on the evolution of the object, in particular
with the indication that the system is more massive 
and young. A general outline of the scenerio presented below (Sect.~\ref{merg_sc}) 
is followed by simple simulations of the merger remnant in Sect.~\ref{merg_sim}.
A discussion of the observations within the merger scenario is given in 
Sect.~\ref{merg_disc}.

\subsection{General considerations and estimates  \label{merg_sc}}

We assume, following Tylenda et al. (\cite{tss}), 
that the progenitor of V838~Mon
was a main sequence star of $M_1 \simeq 8\,M_{\sun}$ or a somewhat less massive
star in the pre-main-sequence phase. In the following we will call it the primary. 
Most likely it forms a wide
binary with a similar star, seen as a B3\,V companion in the decline phase of V838~Mon.
It is possible that the system also had another component, or 
other components, of much lower mass.
Interactions within the multiple system might have destabilizated the orbit
of one of the low mass components causing it to enter a highly eccentric orbit and,
finally, to interact with the primary.
As the whole system is rather young, as discussed in Tylenda et al. (\cite{tss}),
another
possibility is that the V838~Mon progenitor and its presumably low mass companion 
are surrounded by a protostellar disc. Interactions between the disc and the 
low mass star might have caused migration of the latter and growth of the orbit eccentricity,
as discussed, for instance, in Artymowicz et~al. (\cite{artym}) for stellar binaries
or in Armitage \& Bonnell
(\cite{armbonn}) for migrating brown dwarfs arround low mass stars,
leading to a merger with 
the primary. In the case of a primary at the end of the main sequence phase,
the onset of fast expansion of the star might
have contributed to triggering the merger process.

The mass of the accreted companion, $M_2$, in V838~Mon can be estimated from the total
energy budget of the event. As discussed in Sect.~\ref{observ}, during 
the main outburst (January -- mid-April~2002) V838~Mon lost 
$\sim 2.5 \times 10^{46}$~ergs in radiation and presumably $0.01 - 0.1\ M_{\sun}$
in mass loss. The energy necessary to lift this matter from the surface
of a star having $M_1 = 8\ M_{\sun}$ and $R_1 = 5\ R_{\sun}$ and to accelerate 
it to 300\kms\ at large distances is $(0.4 - 4.0) \times 10^{47}$~ergs.
Tylenda (\cite{tyl05}) estimated
that the mass of the inflated envelope contracting in the decline phase was
$M_\mathrm{env} \simeq 0.2\ M_{\sun}$. This value was obtained assuming 
a $n = 3/2$ polytropic model of the envelope. 
If a $n = 3$ polytropic model is assumed, as discussed below,
then the result would be $M_\mathrm{env} \simeq 0.1\ M_{\sun}$. The energy stored
in such an envelope, i.e. a difference between the values obtained from
Eq.~(A.11) in Tylenda (\cite{tyl05}) putting $R_\mathrm{env} = 2000\ R_{\sun}$
and $R_\mathrm{env} = R_1$, is $(2.5 - 6.0) \times 10^{47}$~ergs.
Thus the total energy involved in the V838~Mon event is 
$(3 - 10) \times 10^{47}$~ergs. Equating this value to $(G M_1 M_2)/(2 R_1)$
one obtaines $M_2 = 0.10 - 0.33\ M_{\sun}$.

The above analysis can be repeated assuming an A-type pre-main-sequence star
progenitor discussed in Tylenda et~al. (\cite{tss}), i.e.
taking $M_1 = 5\ M_{\sun}$ and $R_1 = 7.5\ R_{\sun}$. In this case $M_\mathrm{env}$
becomes $\sim 0.3\ M_{\sun}$ and $\sim 0.15\ M_{\sun}$ for the $n = 1.5$ and
$n = 3$ polytropes, respectively. The mass of the accreted component is then 
obtained as $M_2 = 0.15 - 0.5\ M_{\sun}$.

As discussed in Tylenda et~al. (\cite{tcgs}) the case of V4332~Sgr 
can be accounted for by a collision of a planet-like component
with a $\sim 1\,M_{\sun}$ main sequence star.

Let us consider a grazing collision, rather
than a headon collision, as more plausible in the case of a binary merger
in an initial eccentric orbit.
The gravitational plus kinetic energy of the binary system before merger is
\begin{equation}
  E_\mathrm{orb} = -\frac {G\,M_1\,M_2}{2\,r_\mathrm{p}} (1-e),
\label{eq:e_orb}
\end{equation}
where $r_\mathrm{p}$ is the orbital separation at periastron and
$e$ is the eccentricity. For a grazing merger we have
$R_1 \la r_\mathrm{p} \la R_1+R_2$ where $R_2$ is the radius of the
secondary. In the following we will use $r_\mathrm{p} \simeq R_1$.

We mostly consider low-mass main-sequence or pre-main-sequence stars as
secondaries. Their structure can be well approximated by a polytropic
star with an index $n = 3/2$. Thus
the total internal (gravitational plus thermal) energy of the secondary
can be written as (see Eq.~\ref{e_pms})
\begin{equation}
  E_2 = - \frac{3}{7} \frac{G\,M_2^2}{R_2}.
\label{eq:e_2}
\end{equation}

It is reasonable to assume that in the case of a grazing collision
the low mass secondary, if it is not significantly denser than the more massive
primary, is disrupted in the merger process
and forms an inflated envelope of the merger remnant. The primary, 
remaining almost undisturbed, forms a core of the remnant. This scenario is
confirmed by the simulations presented in Sect.~{\ref{merg_sim}. The structure
of the remnant envelope is determined by the distribution of the energy dissipated
in the merger process. The latter is a complex process as demonstrated in
numerical simulations (e.g. Lombardi et~al. \cite{lomb96}, \cite{lomb02}). 
However, to a first approximation
we can assume that the rate of the energy dissipation is 
proportional to the density. In this case the structure of the remnant envelope
can be approximated by a uniform energy source model, 
which is close to an $n = 3$ polytrope (see e.g. Cox \& Guili \cite{cox}).
Indeed, the density distribution in the remnant envelopes obtained in 
our simulations discribed in Sect.~{\ref{merg_sim} (not only in the cases shown
in Fig. \ref{mmas_fig}) can be well reproduced by 
the $n = 3$ polytropic distributions (see Eq. A4 in Tylenda \cite{tyl05}).
For an outer radius of the envelope, $R_\mathrm{env}$,
much larger than the inner one (assumed to be the radius of the
primary, $R_1$) the total energy of the envelope, $E_\mathrm{env}$, 
can be approximated by (see Eq.~A.17 in Tylenda \cite{tyl05})
\begin{equation}
  E_\mathrm{env} \simeq -\frac {G\,M_1\,M_2}{2\,R_1} 
        \frac{1}{\ln(R_\mathrm{env}/R_1)}.
\label{eq:e_env}
\end{equation}

The energy balance can be written as
\begin{equation}
  E_\mathrm{orb} + E_2 = E_\mathrm{env},
\label{eq:balance}
\end{equation}
which, using Eqs.~(\ref{eq:e_orb}) -- (\ref{eq:e_env}), can be transformed to
\begin{equation}
  \ln \frac{R_\mathrm{env}}{R_1} \simeq \left[ (1 - e) + 
   \frac{6}{7} \frac{M_2}{M_1} \frac{R_1}{R_2} \right]^{-1}.
\label{eq:r_env}
\end{equation}

In the above energy balance we do not consider energy losses in form of radiation
and mass loss. In the global energetics of the event they can be important,
as shown in our estimates above for V838~Mon. However, these losses
mostly take place in later phases of the evolution of the merger remnant, which
occur on a thermal time scale. The merger is expected to occur on a dynamical time
scale so for the initial structure of the merger remnant the losses are not
expected to be important. This is consistent with
numerical simulations of stellar mergers (e.g. Lombardi
et~al. \cite{lomb02}, see also Sect.~\ref{merg_sim}).

Analyses of observations give $R_\mathrm{env}/R_1 \gg 1$. Maximum values of the 
photospheric radius in the case of V838~Mon reached $\sim 3000\,R_{\sun}$ 
(Tylenda \cite{tyl05}). This however refers to freely expanding matter rather
than to a hydrostatic remnant. The latter was smaller but probably still as large
as $\sim 2000\,R_{\sun}$, as inferred from the photospheric radius observed 
during the early decline. For $R_1 = 5\,R_{\sun}$ this corresponds to 
$R_\mathrm{env}/R_1 \simeq 400$. In the case of V4332~Sgr the envelope
was probably less inflated but still as large as 
$R_\mathrm{env}/R_1 \simeq 100-150$ (Tylenda et~al. \cite{tcgs}).

Eq.~(\ref{eq:r_env}) shows that from a merger of two similar stars, e.g. two
main sequence stars of not vary different mass, we do not expect
a significantly inflated remnant. For instance, taking $M_1 = M_2$ and $R_1 = R_2$,
Eq.~(\ref{eq:r_env}) predicts $1.7 \la R_\mathrm{env}/R_1 \la 3.2$ for
$0 \le e \le 1$. Note, however, that the cases of two similar stars are not particularly
applicable to the above approach, which assumes an undisturbed primary and a largely
inflated remnant envelope. Simulations, described in Sect.~\ref{merg_sim}
for two $0.6\,M_{\sun}$ ZAMS stars give $6.5 < R_\mathrm{env}/R_1 < 8.4$ for
$0.5 < r_\mathrm{p}/(R_1 + R_2) < 1.0$, thus the remnants are more extended than 
predicted from Eq.~(\ref{eq:r_env}) but still much less than observed.

Taking $M_1 = 8\,M_{\sun}$, $M_2 = 0.3(0.1)\,M_{\sun}$, i.e. masses expected
for V838~Mon, and the corresponding main sequence
radii, i.e. $R_1 = 5\,R_{\sun}$ and $R_2 = 0.35(0.15)\,R_{\sun}$, 
one still gets little
expansion from Eq.~(\ref{eq:r_env}), i.e. $R_\mathrm{env}/R_1 \la 9.0(16.5)$.
However the above set of stellar parameters is not realistic. From
evolutionary considerations it can be concluded that at 
an age of an $8\,M_\odot$ star leaving the main sequence, 
stars of masses $\la 1.0\,M_{\sun}$ still have to be in the pre-main-sequence phase.
As can be seen from Eq.~(\ref{eq:r_env}), larger values of $R_2$ would result 
in more inflated merger remnants.
An $8\,M_{\sun}$ star reachs the ZAMS at an age of $3.1 \times 10^5$~years
(see Sect.~\ref{merg_sim}).
At this age a $0.3(0.1)\,M_{\sun}$ pre-main-sequence star is expected to have a radius
of $\sim 2.5(1.3)\,R_{\sun}$, as estimated from Eq.~(\ref{r_pms}).
To get envelope extensions as large as $R_\mathrm{env}/R_1 \simeq 400$ 
Eq.~(\ref{eq:r_env}) requires $R_2 \ga 0.95(0.32)\,R_{\sun}$.
 
If the V838~Mon progenitor was an early A-type pre-main-sequence star,
we can take $M_1 \simeq 5\,M_{\sun}$ and $R_1 \simeq 7.5\,R_{\sun}$. 
The age of the star would be $\sim 5 \times 10^5$~years (see Sect. \ref{merg_sim}).
At this age the radius of a $0.5(0.15)\ M_{\sun}$ pre-main-sequence star would be 
$\sim 3.5(1.5)\ R_{\sun}$. For these parameters (and assuming $e = 1$) 
Eq.~(\ref{eq:r_env}) predicts $R_\mathrm{env}/R_1 \simeq 230(2400)$.

In the case of V4332~Sgr we can assume
$M_1 = 1\,M_{\sun}$, $R_1 = 1\,R_{\sun}$ and $R_2 = 0.1\,R_{\sun}$ (radius typical
for brown dwarfs and Jupiter-like planets). Then inflations of
$R_\mathrm{env}/R_1 \simeq 150$ can be obtained from Eq.~(\ref{eq:r_env}) if
$M_2 \la 0.02\,M_{\sun}$.

Even in the most favorable cases, i.e. when the term
$(M_2/M_1)(R_1/R_2)$ in Eq.~(\ref{eq:r_env}) is negligible,
$e \ga 0.8$ is required to get the remnant extension as large as 
$R_\mathrm{env}/R_1 \ga 150$.

\subsection{Numerical simulations \label{merg_sim}}

We have carried out simulations of a merger of two stars with parameters expected
for V838~Mon. We used the MMAS (version 1.6)
package described in Lombari et~al. (\cite{lomb02}, \cite{lomb03}) and available
at http:/faculty.vassar.edu/lombardi/mmas/. This code, based on simple physical
arguments and algorithms, allows one to obtained a one dimensional structure
of the collision product without hydrodynamic simulations. It still accounts for
shock heating, mass loss and fluid mixing.
The remnants returned by MMAS are in hydrostatic equilibrium but before thermal
relaxation. The code explicitely assumes $e = 1.0$ before collision so its
results can be considered as reasonable for highly eccentric binary orbits.

Following the discussion in Sect.~\ref{merg_sc}
we have considered an $8.0\,M_{\sun}$ main sequence primary which 
accretes an $0.3\,M_{\sun}$ pre-main-sequence secondary.
As argued in Sect.~\ref{merg_sc}, a grazing merger
is more likely than a headon collision, so we discuss cases with
$0.5 < x_\mathrm{p} \le 1.0$, where $x_\mathrm{p} = r_\mathrm{p}/(R_1 + R_2)$. 

The structure of the primary before
merger was obtained from the TYCHO (version 6.0) stellar evolution code
(Young et~al. \cite{young-arnett})
available at http://chandra.as.arizona.edu/$\sim$dave/tycho-intro.html.
Starting from an initial (pre-main-sequence) model the $8.0\,M_{\sun}$ star reaches 
the ZAMS at an age of $3.15 \times 10^5$~years. For our simulations
we have taken the stellar structure 
at $5.7 \times 10^6$~years. The star
is well on the main sequence with
$\sim 10\%$ of hydrogen burnt in the core and an outer radius $R_1 = 3.7\,R_{\sun}$.

The structure of the secondary has been approximated by a $n = 3/2$ polytrope,
as discussed in Appendix~\ref{app}.
Using Eq.~(\ref{r_pms}) one gets
that at an age of $5.7 \times 10^6$~years
the radius of the $0.3\,M_{\sun}$ secondary is 
$R_2 \simeq 0.9\,R_{\sun}$.

Density profiles
for the stars (before merger) modelled as explained above are shown 
with dashed curves in Fig.~\ref{mmas_fig}.
Full curves present the density profile of the merger products
obtained with MMAS. 
Three cases of $x_\mathrm{p} = 0.6$, 0.8 and 1.0 are shown.
We have defined a parameter to which values of $-8$ and $-12$ have been assigned
for the matter in the primary and secondary, respectively, before merger. MMAS sorts
and mixes
the matter from both components after the merger event so the returned value of this
parameter allows us to see from which component and in what proportion 
the matter comes 
at a given radius of the remnant. This result is shown with dotted curves 
in Fig.~\ref{mmas_fig}.

\begin{figure}
\centering
  \resizebox{\hsize}{!}{\includegraphics{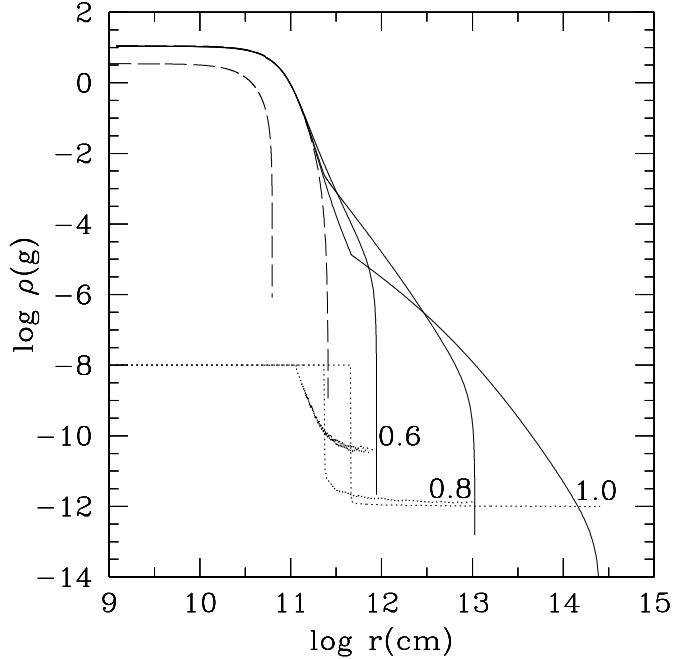}}
  \caption{Results of a grazing collision of two stars using the MMAS package.
The initial system consists
of an $8.0\ M_{\sun}$ main sequence primary and an $0.3\ M_{\sun}$ pre-main sequence
secondary at an evolutionary age of $5.7 \times 10^6$~years. 
Dashed curves: density profiles of the stars (upper: primary -- 
lower: secondary) before collision. Full curves:
density profile in the merger product. Dotted curves: values of a parameter 
showing the origin of the matter in the merger product ($-8$: matter from 
the primary, $-12$: matter from the secondary -- see text). 
The curves (full and dotted) 
are marked with the value of $x_\mathrm{p} = r_\mathrm{p}/(R_1 + R_2)$.
}
  \label{mmas_fig}
\end{figure}

In the case of a main sequence star the density gradient is relatively flat 
in the interior, that
quickly drops when approaching the star surface. Therefore if the star suffers
a grazing collision with a low mass pre-main sequence star its interior structure
remains almost intact 
(see Fig.~\ref{mmas_fig}). Only outer regions are seriously 
disturbed and together with the matter from the destroyed secondary 
they form an inflated
envelope.\footnote{The situation is qualitatively different for collisions
approaching headon, i.e. $x_\mathrm{p} \la 0.4$. Then most of the secondary
dives into the primary centre and forms the core of the collision product.}
The outer radius of the envelope is very sensitive to $x_\mathrm{p}$ and
it becomes as large as $3.7 \times 10^3\,R_{\sun}$ in the case of $x_\mathrm{p} = 1.0$
shown in Fig.~\ref{mmas_fig}.
Note that Eq.~(\ref{eq:r_env}) predicts $R_\mathrm{env} \simeq 7 \times 10^3\,R_{\sun}$
which is in a fair agreement with the simulations.
The mass of the inflated envelope is $\sim 0.30\,M_{\sun}$.
For $x_\mathrm{p} \ga 0.8$
the remnant envelope is predominantly formed from the disrupted secondary. 
The simulations predict $\sim 0.015\,M_{\sun}$ lost from the system, 
for $x_\mathrm{p} \ga 0.8$ almost entirely from the secondary.

We have also carried out simulations for the same stars as in Fig.~\ref{mmas_fig}
but at an age of $3 \times 10^7$~years as well as for 
two pre-main-sequence stars of 5.0 and $0.3\ M_{\sun}$ at an age of
$5 \times 10^5$~years. In the former case
the $8\ M_{\sun}$ primary just started 
to move from the main sequence and the radii of the components were 
$R_1 = 9\,R_{\sun}$ and $R_2 = 0.6\,R_{\sun}$.
In the latter case the effective temperature
of the $5\ M_{\sun}$ primary was $\sim$~10\,000~K, 
i.e. corresponding to an $\sim$~A0 spectral type discussed 
in Tylenda et~al. (\cite{tss}). The stellar radii were
$R_1 = 8\,R_{\sun}$ and $R_2 = 2\,R_{\sun}$. The results were very similar to
those displayed in Fig.~\ref{mmas_fig}. In particular the radii of 
the obtained remnants were $10^2-10^4\ R_{\sun}$.

In all the cases discussed the density
distribution in the inflated envelope of the remnant can be well approximated
by a polytropic envelope (see Appendix in Tylenda \cite{tyl05}) with an index
$n = 3$.

\subsection{Discussion  \label{merg_disc}}

As we have shown using simple estimates in Sect.~\ref{merg_sc} and simulations
in Sect.~\ref{merg_sim}, it is possible to obtain merger remnants with radii similar 
to those observed in V838~Mon and V4332~Sgr. These results should be verified with 
more sophysticated modelling. As noted in Sect.~\ref{merg_sim} the MMAS code
is a simple algorithm. It has been calibrated on SPH simulations of mergers of 
two low-mass ($0.4 - 0.8\ M_{\sun}$) main-sequence stars. 
For the star parameters used in
Sect.~\ref{merg_sim} the code works at its limits and the results are probably 
subject to considerable uncertainties. 

Some important processes are not taken into account in the MMAS code.
These include effects of radiation transfer and angular momentum.
Radiation pressure becomes particularly important if the luminosity
is super-Eddington. This was certainly the case in V838~Mon. 
In February--March~2002 the observed luminosity was above 
the Eddington value (Tylenda \cite{tyl05}), which means that 
the rate of energy dissipation deep in the
envelope must have been super-Eddington by a larger factor. A part of energy
went to expansion of the envelope and mass loss.
Also energy diffusion processes in the envelope work in the sense that 
the observed eruption lasts longer, while the maximum luminosity is lower, 
compared to the original energy burst. 
Indeed, an analysis done in Tylenda (\cite{tyl05})
suggests that the main eruption observed in February-March~2002 was generated
on a time scale of a few days at the end of January~2002.

The rate of energy dissipation in a merger event can be estimated from
\begin{equation}
  L_\mathrm{diss} \simeq \frac{1}{2}\ \rho\ v^3\ \pi\ R_2^2,
\label{eq:ldiss_1}
\end{equation}
where $v$ is the velocity of the accreted component of radius $R_2$ moving
in an ambient medium of density $\rho$. Assuming that the merger takes place
in the outer layers of the primary and that $v$ is close to the Keplerian velocity, 
i.e. $v \simeq \sqrt{G M_1/R_1}$, Eq.~(\ref{eq:ldiss_1}) becomes
\begin{eqnarray}
\label{eq:ldiss_2}
  L_\mathrm{diss} & \simeq & 3 \times 10^{10}\ L_{\sun}
     \left( \frac{\rho}{0.1\ \mathrm{g}\ \mathrm{cm}^{-3}} \right)
     \left( \frac{M_1}{8\ M_{\sun}} \right)^{3/2}
\nonumber \\
 & & \left( \frac{R_1}{5\ R_{\sun}} \right)^{-3/2}
     \left( \frac{R_2}{1\ R_{\sun}} \right)^2.
\end{eqnarray}
This is a very rough estimate but shows that in the case of mergers simulated in
Sect.~\ref{merg_sim} the energy is expected to be dissipated at a very high
rate, orders of magnitude above the Eddington limit. This can have important effects
on the structure of the remnant and mass loss. In particular the total mass
lost in the merger can be significantly higher than the values predicted
from dynamical effects only in the MMAS code ($0.01-0.02\ M_{\sun}$ in our cases).

The evolution of a merger remnant can be quite easily predicted, 
at least qualitatively. The high energy sources 
will quickly decline after the merger.
The rate of the energy transported from the base of 
the inflated envelope will soon become limited to the luminosity of the
central star, which is much less than the luminosity radiated away 
from the photosphere. As a result, the matter in the envelope, which did not achieve
escape velocity during the merger eruption, has to contract under 
the gravity of the central star. In this way
the internal energy of the contracting envelope can be liberated 
at a rate necessary to balance the photospheric energy loss.

A transition between the eruption phase (dominated by expansion of the matter)
and the decline phase (dominated by gravitational contraction of the envelope)
is expected to occur on a dynamical time scale, which is
\begin{equation}
  t_\mathrm{d} \simeq \left( \frac{R_\mathrm{env}^3}{2\ G\ M_1} \right)^{1/2}.
\label{eq:td}
\end{equation}
For the observed radius of V838~Mon at the beginning of the decline,
$R_\mathrm{env} \simeq 2000\ R_{\sun}$, (and assuming $M_1 = 8\ M_{\sun}$)
Eq.~(\ref{eq:td}) gives $t_\mathrm{d} \simeq 1$~year. Taking 
$R_\mathrm{env} \simeq 150\ R_{\sun}$ and $M_1 = 1\ M_{\sun}$
as the parameters for V4332~Sgr (Tylenda et~al. \cite{tcgs}) one gets
$t_\mathrm{d} \simeq 1$~month. During this 
transition phase the photospheric regions are deprived of energy supply
(merger energy is declining, contraction not yet effectively started)
so they must cool down due to radiative processes. In V838~Mon this phase was 
observed over $\sim 200$~days since $\sim 10$~April~2002.
As can be seen from Fig.~2 in Tylenda (\cite{tyl05}), the object kept
the effective radius between $2000-3000\ R_{\sun}$, 
its effective temperature
dropped from $\sim 3600$~K to a minimum value of $\sim 1800$~K, while
the luminosity declined by more than an order of magnitude.
In the case of V4332~Sgr the transition phase probably started on 
$\sim 5$~April~1994 (Tylenda et~al. \cite{tcgs}). 
During the subsequent three months 
the object dropped by a factor of $\sim 100$ in luminosity, reaching an effective
temperature as low as $\sim 2300$~K (Martini et~al. \cite{martini}).

In later phases the evolution of a merger remnant will
be governed by the gravitational contraction of its inflated envelope.
Thus the main processes are here the same as in protostars. As a result 
we expect that on the HR diagram the object should decline along 
a Hayashi track.
However the time scale of evolution of the merger remnant must be much shorter 
than that of a protostar, as instead of the whole star, 
only a small mass envelope is involved in the contraction. As shown in Tylenda
(\cite{tyl05}) and Tylenda et~al. (\cite{tcgs}) the observed evolution
of V838~Mon for $\sim 2.5$~years after its 2002 eruption and the observed
state of V4332~Sgr $\sim 9$~years after its 1994 eruption are
consistent with this prediction.

This behaviour of V838 Mon type objects
in initial decline resemble the expected evolution of supernova type Ia
companions as discussed in Podsiadlowski (\cite{podsiadlo}). These stars,
when puffed up to large radii and luminosities by the supernova explosion, 
due to a very short thermal time scale
are expected to enter into the "forbidden" Hayashi region and then decline rapidly.

Angular momentum is not taken into account 
when determining the remnant structure in the MMAS code.
The code however provides the expected angular 
momentum distribution in the remnant. In our cases the ratio of the resultant 
specific angular momentum to the local Keplerian value is roughly constant 
throughout the inflated envelope and is typically 0.10--0.15.
This may have important effects on the structure of
the remnant, particulary on its evolution with time. The above result means
differential rotation of the envelope with the rotation velocity increasing inward
as $\sim r^{-1/2}$. Radial motions, for instance convection, would transport
angular momentum outward, thus accelerating rotation in outer layers. This may lead
to an expansion of the envelope and/or intensified winds from the remnant 
in the equatorial directions. In later epochs, when the envelope as a whole 
is expected to contract, the angular momentum may prevent equatorial regions from
significant contraction. Thus some matter can be left at 
larger distances, forming a Keplerian ring or disc arround the contracting central
object. As discussed in Tylenda et~al. (\cite{tcgs}) this is a likely explanation
of the origin of cold matter seen in the emission molecular bands and 
atomic lines in the spectrum of V4332~Sgr.

One of the principal observational characteristics of an eruptive object 
is its light curve. 
V838 Mon displayed a quite complex light curve in January--April~2002,
which can be interpreted as a sequence of three consecutive outbursts. This led
Retter \& Marom (\cite{ret1}) to propose that the V838~Mon eruption was due to
a red giant that swallowed three planets. This idea however suffers from several
important shortcomings. First, as shown in Tylenda et~al. (\cite{tss}) 
the progenitor of V838~Mon was not a red giant. Second, the energy budget
of the V838~Mon eruption in Sect.~\ref{merg_sc} is such that a mass two orders
of magnitude larger than three Jupiter-like planets is required to account for it.
Third, it seems impossible to propose a scenario in which three planets
would fall onto the central star one by one on a time scale of months.
A study in Tylenda (\cite{tyl05}) shows that if a difference
in propagation time between the second (observed in February~2002) and 
the third (observed in March~2002) outburst is taken into account
then the conclusion is that the two outbursts were generated on
a time span of a few days only.
The whole outburst can be divided into two main phases.
The first one, observed in January~2002 and called 
pre-eruption in Tylenda (\cite{tyl05}), was marked by a relatively static photosphere
of $\sim 350\ R_{\sun}$. The second one was observed in February--April~2002,
called eruption, and was marked by an
expansion of the photosphere from $\sim 350$ to $\sim 3000\ R_{\sun}$.

The question that arises is how this multi-outburst light curve of 
V838~Mon can be
explained in our merger model. Two general interpretations can be
proposed to account for it. The first one is that the event involved accretion 
of two companions on a time span of a month. The second is that it was a single
companion that collided and merged with the V838~Mon progenitor but that the
merger event proceeded in multiple phases.

Accretion of two companions can occur in two scenarios.
One possibility is that orbits of two low-mass companions had been disturbed, 
say due to a destabilizing event in the V838~Mon system,
in such a way that the two companions collided with the
primary star one after the other. Another
possibility is that only one companion had been disturbed in its orbit so
it collided with the primary but the merger event destibilized another
companion causing its merger with the primary a month after the first one.

A more detailed scenario in the former case is as follows.
The first companion collided with the primary star at the end of December~2001 
and initiated the pre-eruption phase. As a results an envelope inflated to
$\sim 350~R_{\sun}$ was formed. If there were another companion orbiting the primary
at a distance much smaller than the radius of the envelope then it would become 
engulfed by the envelope. 
Drag forces between the orbiting companion and the envelope caused the companion 
to spiral in, so after a month it reached dense regions of the primary, more or less
undisturbed by the accretion of the first companion. 
Then the second companion was disrupted,
liberating its orbital energy which resulted in the main eruption started at the
begining of February~2002. However, as we show below, it is very difficult,
perhaps even impossible,
to get a spiraling time as short as 1~month for realistic parameters in this
scenario. 

The luminosity of V838~Mon in the main eruption
was $\sim 20$ times higher than that in the pre-eruption. 
This implies that the second
companion was more massive than the first one by a factor similar to the
luminosity ratio. Thus we can assume that the mass of the first companion 
was $\sim 0.01\ M_{\sun}$.
Let us approximate the structure of the envelope resulting from its disruption
by a polytropic model (see Appendix in Tylenda \cite{tyl05}).
The radius of the envelope was $\sim 350\ R_{\sun}$, as observed in the pre-eruption phase,
while its mass can be assumed to be comparable
to the mass of the distrupted companion. Taking the mass and radius of the
primary, $M_1 = 8\ M_{\sun}$ and $R_1 = 5\ R_{\sun}$, we can estimate 
a density at the base
of the envelope, $\rho_0$, from Eq.~(A.5) in Tylenda (\cite{tyl05}). In the
case of the polytropic index $n = 3/2$ one gets 
$\rho_0 \simeq 3 \times 10^{-7}$\g\cc, while for $n = 3$ the result is
$\rho_0 \simeq 1.5 \times 10^{-5}$\g\cc. Now we can use Eq.~(\ref{ts_spir})
to estimate the time necessary for the second companion to spiral in, 
$t_\mathrm{s}$.
Assuming $M_2 \simeq 0.3\ M_{\sun}$, $R_2 \simeq 0.9\ R_{\sun}$ 
(as in Sect.~\ref{merg_sim})
and taking a rather lower limit to the orbital separation of $\sim 2\,R_1$, i.e.
putting $r_0/R_0 \simeq 2.0$ in Eq.~(\ref{ts_spir}), the result is 
$t_\mathrm{s} \simeq 150$ and 6~years for $n = 3/2$ and 3, respectively.
As can be seen from Eq.~(\ref{ts_spir}) $t_\mathrm{s}$ strongly increases
with increasing values of $r_0/R_0$, e.g. for $r_0/R_0 = 10$ the in-spiral
time becomes 5000 and 2000~years, respectively.

Another point is that during the spiraling phase the companion releases energy
$\sim [G M_1 M_2 (1 - R_0/r_0)]/(2 R_0)$ which for $R_0/r_0 \ll 1$ becomes
comparable to the energy released during the disruption of the companion
at $R_0$. Thus even if it were possible to spiral in the companion over
a month, the energy released during the pre-eruption phase would be comparable
to that in the eruption phase. In the case of V838~Mon the ratio of the energy
radiated away during the pre-eruption phase (January~2002) to that in the
eruption phase (February -- mid-April~2002) was $\sim 0.03$.

Therefore we can conclude that if the 2002 eruption of V838~Mon was due to
accretion of two star-like objects the only possibility seems to
be that the two bodies arrived from outside on very similar trajectories: 
the lower-mass one was followed by 
the more-massive one with a month between them.
It could have been a low-mass binary (e.g. a $\sim 0.3\ M_{\sun}$ star orbited by 
a $\sim 0.01\ M_{\sun}$ brown dwarf) "injected" by dynamical effects from 
the periphery of the V838~Mon system or from outside it.

In our opinion, the second interpretation is more reasonable.
The outburst was likely due to the merger of a single companion but that the
merger proceeded in multiple phases. 
In several numerical simulations of star
collisions described in the literature this is indeed the case (e.g.
Lombardi et al. \cite{lomb96}, Freitag \& Benz \cite{freibenz},
see also animations available at http://www.manybody.org/modest).
The components, often partly disrupted after the first encounter, 
leave each other on a highly eccentric orbit
and collide again at the return. This may be repeated, sometimes several
times, with a rapidly decreasing time between the encounters.
Thus a plausible interpretion of the light curve of V838~Mon is as follows.
The first major encounter betweeen the low-mass companion and the progeniotor
of V838~Mon happened at the end of December~2001 and produced the pre-eruption
phase. The companion probably became partly disrupted and only a small part of the
available energy was dissipated. Matter significantly shocked during the first
encounter formed the pre-eruption envelope. Most of the matter from the companion
collided with the primary after a month (note that the dynamical time scale 
-- Eq.~(\ref{eq:td}) -- for the observed
size of the envelope in the pre-eruption is $\sim 30$~days)
and once again after a few days. 
This produced the main eruption with the double-maximum structure
in the light curve observed in February and March~2002 and two shells in mass loss
found by Tylenda (\cite{tyl05}).

Before the main encounters, which led to the observed outburst, 
it is possible that there had been a series of minor encounters
due to the companion passing by the periastron. This could have led to
minor outbursts and a long-term increase in brightness before the main
outburst. In the case of V838~Mon we have no observational data between 1994 and
the outburst discovered at the beginning of 2002. 
Before 1994 the object had remained constant, at least
for $\sim$55~years (Goranskij et~al. \cite{goran}). In the case of V4332~Sgr,
however, Kimeswenger (\cite{kimes05}) shows that the object started to rise
several years before the outburst observed in 1994. This can be interpreted
as a result of minor interactions before the final merger. If V4332~Sgr 
is a Galactic buldge object, then its progenitor
was a G-type giant, as discussed in Tylenda et~al (\cite{tcgs}).
Then the merger process may have occured well below the giant photosphere. 
Before the merger, when
the companion was spiraling in the giant envelope, the brightness of the object
was expected to steadily increase, as can be seen from Eq.~(\ref{lum_spir}).

We now discuss the observed abundances in the V838~Mon type objects.
As discussed in Sect.~\ref{observ}, there is no observational indication 
that the matter observed in the objects
was significantly processed by nuclear burning. 
In our merger model the observed matter is primarily due to
the disrupted low-mass secondary. In the case of V838~Mon we argue that it was
a low-mass pre-main-sequence star so no significant nuclear processing took
place inside it. In particular this concerns the observed abundance of Li.
Low-mass stars of masses $\la 0.5\ M_{\sun}$ are not expected 
to have significantly burnt Li at an age of 
$\la 10^7$~years (Chabrier \& Baraffe \cite{chabar}, also our own modelling
done with the TYCHO code). Even if some mass from the primary had been mixed into
the remnant it should have no significant effect on the surface Li abundances.
Li is expected to remain unburnt in outer layers of an $8\ M_{\sun}$ main-sequence
or a $5\ M_{\sun}$ pre-main-sequence star.

\section{Summary and discussion \label{summary}}

\begin{table*}
\begin{minipage}[t]{\columnwidth}
\caption{V838 Mon versus models}
\label{comp_tab}
\centering
\renewcommand{\footnoterule}{ }  
\begin{tabular}{llll}
\hline \hline
Observed property  & Classical nova  & Born-again AGB   & Stellar merger  \\
~~~~~~~~~(1)       & ~~~~~~~~(2)     & ~~~~~~~(3)       & ~~~~~~~~(4) \\
\hline \hline
Increase by factor $\sim 10^3$ & (+)\footnote{(+) can be explained by the model; 
($-$) cannot (very difficult to) be explained by the model} 
  Possible, although & (+) Possible if starts well 
                           & (+) Easy to obtain \\
in luminosity ($\Delta V \simeq 7.5$)& usually larger & down the cooling track \\ 
\hline
Multi-outburst  & (+) Compatible with  & ($-$) Cannot be explained  
                                      & (+) Multiple phase  \\
light curve & a slow nova  & (Lawlor's accretion episode  & merger process\\
       &  & cannot work) & \\
\hline
Fading as a very & ($-$) Contrary to expected  & ($-$) Contrary to expected
                                          & (+) Contraction along\\
cool supergiant & (decline at $T_\mathrm{eff} \ga 10^5$~K) 
                & (decline at $T_\mathrm{eff} \ga 10^5$~K) 
                                    & the Hayashi track \\
\hline
$L\simeq 10^6 L_{\sun}$ & ($-$) Too luminous for a slow &($-$) Too luminous
                                      &  (+) Merger of $\sim 0.3M_{\sun}$ \\
 over $\sim 70~$days & nova, too long for a fast nova  & and too fast
                                     & and $\sim 8M_{\sun}$ stars  \\
\hline
$\la$ solar abundances  & ($-$) $\sim 10 \times$solar abundances 
                               & ($-$) Enrichment by & (+) Young stars at\\
($[Fe/H] \simeq -0.3 $) & expected & dredge-up expected & galactic outskirts \\
\hline
Outflow velocity & ($+$) Observed in slow novae &($-$) Too fast 
 & (+) $\sim$ escape velocity \\
$200-500 \km \s^{-1}$   &  & & from a MS star\\
\hline
B-type progenitor & ($-$) Cannot work  & ($-$) Too cool for a star
                                          & (+) B-type MS primary\\
                &            & before born-again AGB  & \\
\hline
Association with         & ($-$) too little time to form
              &($-$) Incompatible with  &(+) Progenitor of similar type \\
a young (B3\,V) star & and cool a WD & a post-AGB progenitor  &\\
\hline
Circumstellar & ($-$) Past eruptions expected & ($-$) Past AGB wind
                                    & (+) Part of an ISM region \\
non-ionized matter & to sweep up the enviroment & expected to be ionized
                                      & \\
\hline
\hline
\end{tabular}
\end{minipage}
\end{table*}
%

Table \ref{comp_tab} summarizes our discussion of the models
compared with the observations. 
For clarity it is restricted to the observed properties
of V838~Mon only. Column (1) lists the principal observational properties
of the object. Comments on how these properties can be explained
in the classical nova, born-again AGB (very late He-shell flash) and 
stellar merger models are given in columns (2) -- (4), respectively.
In the table we do not discuss the idea of an evolved very massive
star proposed in Munari et~al. (\cite{muna05}). This is mainly because
this idea is ruled out by the observed lack of noticeable ionization 
of the matter seen in the light echo (Tylenda et~al. \cite{tss}). 
There is no available 
model based on this idea so it is not clear what observational properties
are predicted from it.

Table \ref{comp_tab} shows that 
presently the merger model
is the most promising one for V838~Mon and the other objects of the same type.
Both classical nova and He-shell flash models seem to be ruled out
by too many contradictions with the observed properties of the objects.

The main drawback of our merger model is that it is based on estimates
and approximate considerations.
The simple simulations presented in Sect.~\ref{merg_sim} only
demonstrate that an inflated envelope can be formed from a merger.
Realistic simulations should provide a three-dimensional
structure of the inflated envelope, its evolution after merger, and
the expected light curve. This would however be a very complex task involving
three-dimensional radiation-hydrodynamics. We hope that this paper
will stimulate research in this direction.

So far we have seen three events of the V838~Mon type: two in our Galaxy,
one in M31. A binary population synthesis done in
Hen, Podsiadlowski \& Eggleton (\cite{hpe}) shows that 1-2 stellar mergers are
expected in our Galaxy per 10~years. This result considers only binaries with 
stellar components having masses
$\ge 0.8\ M_{\sun}$. Thus if multiple systems, lower mass stars, brown dwarfs 
and massive planets were also considered, two events in our Galaxy in a time span 
of 8~years are not surprising. What is suprising is that no event of this kind 
had been reported (in our Galaxy) before the discovery of V4332~Sgr. 
Kato (\cite{kato}) suggests
that Nova CK~Vul 1670 might have been a stellar merger event. 
Bond \& Siegel (\cite{bs05}) mention that Nova V1148~Sgr 1943 had a late-type
spectrum thus being a possible object of the V838~Mon type.
It is quite probable that some V838~Mon type objects remain hidden among stars
catalogued as classical nova or nova-like. 

Mergers of massive stars are considered
as a possible way to form very high mass stars. Bally \& Zinnecker
(\cite{ballzinn}) have discussed expected observational appearances of mergers 
involving massive stars. These events can release up to $10^{51}$ ergs and 
reach a peak luminosity above $10^7\ L_{\sun}$. They argue that
an event of this kind happened $\sim 500$~years ago in the Orion molecular cloud.
Their study as well as our work shows that stellar mergers are important not only
to understand the evolution of binaries or the origin of blue stragglers. Due to
the large amount of energy released on a short time scale they can also produce 
spectacular events, easy to observe even beyond our Galaxy.

\appendix
\section{Low mass pre-main-sequence stars  \label{app}}

Low mass pre-main-sequence stars are convective. 
In this case, for given values of the star mass, $M_\ast$, and radius, $R_\ast$,
the stellar structure can be obtained from a solution of the Lane-Emden
equation with a polytropic index $n = 3/2$ (for detailes
see e.g. Cox \& Giuli \cite{cox}). 

A pre-main-sequence star radiates mainly due to its gravitational contraction, so its
lifetime, $t_\mathrm{pms}$, can be estimated from (see e.g. Stahler \cite{stahler})
\begin{equation}
   t_\mathrm{pms} \simeq \frac{-E_\ast}{L_\ast}.
\label{t_pms}
\end{equation}
Here $E_\ast$ is the total (gravitational plus internal) energy of the star, which,
for an $n = 3/2$ polytrope, is (see e.g. Cox \& Giuli \cite{cox})
\begin{equation}
  E_\ast = - \frac{3}{7} \frac{G\,M_\ast^2}{R_\ast},
\label{e_pms}
\end{equation}
while the luminosity, $L_\ast$, is
\begin{equation}
  L_\ast = 4\,\pi\,R_\ast^2\,\sigma\,T_\ast^4,
\label{l_pms}
\end{equation}
where $T_\ast$ is the stellar effective temperature.
Combining the above equations one gets
\begin{eqnarray}
\label{r_pms}
 R_\ast & = & f_\mathrm{pms} \left( \frac{3\,G\,M_\ast^2}
        {28\,\pi\,\sigma\,T_\ast^4\,t_\mathrm{pms}} \right)^{1/3} 
\nonumber \\
  & = & 2.56\,f_\mathrm{pms} \left( \frac{M_\ast}{0.3\,M_{\sun}} \right)^{2/3}
         \left( \frac{T_\ast}{3000\,\mbox{K}} \right)^{-4/3} 
\nonumber \\
& &  \left( \frac{t_\mathrm{pms}}{10^6\,\mbox{yrs}} \right)^{-1/3} R_{\sun},
\end{eqnarray}
where $f_\mathrm{pms}$ accounts for an uncertainty due to Eq.~(\ref{t_pms}).
Comparing results of evolutionary model calculations for low mass 
pre-main-sequence stars (Stahler \cite{stahler}, our own models calculated
with TYCHO) with the predictions from Eq.~(\ref{r_pms}) we have found
$f_\mathrm{pms} = 0.6 - 0.8$.

\section{Spiraling in a polytropic envelope  \label{spiral}}

Let us consider a companion of mass $M_2$ and radius $R_2$ 
that orbits inside a non-rotating envelope of a primary star of mass, $M_1$,
at a distance, $r$, from the star centre.
We assume that $M_2 \ll M_1$ and that the density in the envelope, $\rho$,
is much lower than the mean density of the companion, 
$\rho_2 = (3M_2)/(4\pi R_2^3)$. Thus we can neglect effects of accretion or 
evapration processes on the mass of the companion. Then we can assume
that the companion moves with a Keplerian velocity, $v = \sqrt{GM_1/r}$, 
so its energy (gravitational plus kinetic) is
\begin{equation}
  E = - \frac{G\,M_1\,M_2}{2\,r}.
\label{e_spir}
\end{equation}
Due to drag forces the companion loses its energy at a rate which we
approximate by
\begin{equation}
  \frac{\ud E}{\ud t} \simeq -\pi\,R_2^2\,\rho\,v\,\frac{v^2}{2}.
\label{de_spir}
\end{equation}
Taking a time derivative of Eq.~(\ref{e_spir}) and substituting it in
Eq.~(\ref{de_spir}) one gets the spiraling rate
\begin{equation}
  \frac{\ud r}{\ud t} = - \frac{\pi\ R_2^2}{M_2}\,\rho\,(G\,M_1\,r)^{1/2}.
\label{dr_spir}
\end{equation}

We assume that the structure of the envelope is given by a polytropic solution
of index $n$ so, apart from the outermost regions of the envelope,
the density distribution can be approximated by (see Eq.~A.4 in
Tylenda (\cite{tyl05}) for $x \ll 1$ and $x_0 \ll 1$)
\begin{equation}
  \rho \simeq \rho_0 \left( \frac{R_0}{r} \right)^n,
\label{dens_spir}
\end{equation}
where $\rho_0$ is the density at the base of the envelope at $r = R_0$.
Then Eq.~(\ref{dr_spir}) can be rewritten as
\begin{equation}
  r^{n+1/2}\ {\ud r} = 
   - \frac{\pi\ R_2^2}{M_2}\,\rho_0\,R_0^n\,(G\,M_1)^{1/2}\,\ud t,
\label{drr_spir}
\end{equation}
which can be integrated giving
\begin{equation}
  r_0^{n+1/2} - r^{n+1/2} =
    \left( n + \frac{1}{2} \right) \frac{\pi\ R_2^2}{M_2}\,\rho_0\,R_0^n\, 
    (G\,M_1)^{1/2}\,t,
\label{r_spir}
\end{equation}
where $r_0$ is the radius of the orbit of the spiraling companion at $t = 0$.
Eq.~(\ref{r_spir}) can be used to obtain the time, $t_\mathrm{s}$, 
necessary for the comapnion to spiral from the initial orbit at $r_0$ 
to the base of the envelope at $r = R_0$, i.e.
\begin{equation}
  t_\mathrm{s} = \frac{t_0}{n + \frac{1}{2}}
      \left[ \left( \frac{r_0}{R_0} \right)^{n+1/2} - 1 \right],
\label{ts_spir}
\end{equation}
where
\begin{equation}
  t_0 = \frac{2}{3\,\pi}\,\frac{R_2}{R_0}\,
        \frac{\rho_2}{\rho_0}\,P_0,
\label{t0_spir}
\end{equation}
where $P_0 = 2 \pi R_0 \sqrt{R_0/(G M_1)}$ is the orbital period at the base
of the envelope. 

Eqs.~(\ref{r_spir}) and Eq.~(\ref{de_spir}) can be used to
obtain the luminosity
generated by the spiriling companion
\begin{eqnarray}
\label{lum_spir}
  L & = & - \frac{\ud E}{\ud t} \\
 & = & \frac{G\,M_1\,M_2}{2\,R_1\,t_0}
     \left[ \frac{t}{t_\mathrm{s}}
     + \left( 1 - \frac{t}{t_\mathrm{s}} \right)
     \left( \frac{r_0}{R_0} \right)^{n+1/2} 
     \right]^{-\frac{n+3/2}{n+1/2}}.
\nonumber
\end{eqnarray}

\acknowledgement{The research reported in this paper has partly been supported
by the Polish State Committee for Scientific Research through 
a grant no. 2~P03D~002~25, as well as by the Israel Science Foundation.
The authors thank to the referee (Ph. Podsiadlowski) for very useful and constructive
comments.}

\end{document}